\documentclass[aps,pra,10pt,twocolumn,showpacs,superscriptaddress,footnoteinbib]{revtex4}
\usepackage{amsmath}
\usepackage{latexsym}
\usepackage{amssymb}
\usepackage{color}
\usepackage{graphicx}
\usepackage{graphics,epstopdf}
\usepackage{soul}
\usepackage[breaklinks]{hyperref}

\begin{document}

\title{Authentication protocol based on collective quantum steering}
\author{Debasis Mondal}
\thanks{cqtdem@nus.edu.sg} 
\affiliation{Centre for Quantum Technologies, National University of Singapore, 3 Science Drive 2, Singapore 117543}
\author{Chandan Datta}
\thanks{chandan@iopb.res.in}
\affiliation{Institute of Physics, Sachivalaya Marg,
Bhubaneswar 751005, Odisha, India.}
\affiliation{Homi Bhabha National Institute, Training School Complex, Anushakti Nagar, Mumbai 400085, India.}
\author{Jaskaran Singh}
\thanks{jaskarasinghnirankari@iisermohali.ac.in}
\affiliation{Department of Physical Sciences, Indian Institute of Science Education and Research (IISER) Mohali,
Sector 81 SAS Nagar, Manauli PO 140306 Punjab India.}
\author{Dagomir Kaszlikowski}
\thanks{phykd@nus.edu.sg} 
\affiliation{Centre for Quantum Technologies, National University of Singapore, 3 Science Drive 2, Singapore 117543}
\affiliation{Department of Physics, National University of Singapore, 2 Science Drive 3, 117542 Singapore, Singapore}
\date{\today}

\newcommand{\ket}[1]{| #1 \rangle}
\newcommand{\bra}[1]{\langle #1 |}
\newcommand{\braket}[2]{\langle #1 | #2 \rangle}
\newcommand{\ketbra}[2]{| #1 \rangle \langle #2 |}
\newcommand{\cosec}{\operatorname{cosec}}

\begin{abstract}
It is well known that certain quantum correlations like quantum steering exhibit a monogamous relationship. In this paper, we exploit the asymmetric nature of quantum steering and show that there exist states which exhibit a polygamous correlation, known as collective correlation [He and Reid, Phys. Rev. Lett. 111, 250403 (2013)], where the state of one party, Alice, can be steered only by the joint effort of the other two parties, Bob and Charlie. As an example, we explicitly single out a particular set of $3$ qubit states which exhibit this polygamous relationship, known as collective steerability. We provide a recipe to identify the complete set of such states. We also provide a possible application of such states to an information theoretic task, termed as quantum key authentication (QKA) protocol. QKA can also be used in conjunction with other well known cryptography protocols to improve their security and we provide one such example with quantum private comparison (QPC).
\end{abstract}
\maketitle

\vskip 0.5in
\section{Introduction}

Quantum mechanical correlations offer many surprises whenever one digs into the theory to understand its nature and differences from the classical world. Some examples include correlations arising from entanglement~\cite{entanglement}, Bell non-locality~\cite{EPR, Bell_theorem, CHSH}, contextuality~\cite{Kochen1967, Graph_cabello_2014}, coherence~\cite{coh1, coh2} and steering~\cite{steering,jeba_steering,debarshi_steering,tanumoy_steering,manik_steering}. Such correlations have the unique and surprising property of being monogamous~\cite{ent_mono,ent_mono2,ent_mono3,nonlocal_mono,Bell_mono_2009,Context_mono_2012,context_mono_nonlocality}. Correlations between certain parties are said to be monogamous if they diminish when shared among more additional parties. A simple example is illustrated by Bell-CHSH inequality~\cite{Bell_mono_2009}: Two parties Alice and Bob share non-local correlations and are able to violate the Bell inequality. If the state of Alice is also entangled with a third party Charlie, the non-local correlations between Alice and Bob diminish as the correlations between Alice and Charlie increase. It is therefore implied that the Bell-CHSH correlations are monogamous. Monogamy of correlations has been extensively studied and has found widespread applications in information theoretic tasks like key distribution~\cite{nonlocality_key_mono, context_key_mono}. 



Quantum steering~\cite{steering} was first introduced for bipartite parties. In this scenario, Bob prepares an entangled state of joint systems $A$ and $B$. He keeps the system $B$ with himself and transmits the system $A$ to Alice. Alice does not trust Bob, but believes that the system sent to her is quantum. Bob's job is to convince Alice that the system he sent indeed belongs to an entangled state and it is possible for him to steer her state. 

While it is well known that steering is monogamous~\cite{steering_mono, steering_mono2,cv_steering_mono1}, a major aspect of it has not been addressed yet. Steering at its core is asymmetric and although it's monogamous from one side, the nature of correlations from the other side is yet to be studied for discrete systems, while a number of works have been directed towards continouos variable systems~\cite{collective_steering,optical_collective_steering,exp_collective_steering}.  

Consider a scenario, where three parties Alice, Bob and Charlie share an entangled state $\rho_{abc}$. From monogamy of steerability~\cite{steering_mono,steering_mono2} if Alice can steer Bob, she cannot steer Charlie and vice-versa. The idea originates from the resource theory of quantum correlations like entanglement~\cite{resource, entanglement}. However, the questions we address here are firstly, how to single out states for which a particular party (say Alice) cannot be steered independently either by Bob or Charlie but rather only if they steer together and secondly, how such states can be used in cryptographic and communication protocols.  

In this paper, we provide a detailed analysis of quantum steering for such a scenario. We identify a set of states for which Alice can share a polygamous relationship with Bob and Charlie, also known as collective steerability of Alice by Bob and Charlie. We lay down the foundation for identifying the complete set of such states. Continous variable versions of such states, offering this collective steering have been found useful in a number of information processing tasks including quantum communication~\cite{poly_comm} and secret sharing~\cite{secret_sharing}. We also show a possible advantage offered by this collective steerability in the security of quantum key distribution (QKD) protocols. To that end we propose a new key distribution protocol, termed as quantum key authentication (QKA), which utilizes these collectively steerable states~\cite{collective_steering}. The protocol helps authenticate the claims of measurement basis of either party by a subsequent measurement of coherence. Our protocol proves advantageous in reference frame alignment, which is an essential part of QKD, and is hard to achieve using classical communication. Furthermore, it can also be used as a standalone QKD protocol.  The protocol is different than the standard QKD protocols and is made semi-device independent by the aforementioned property of collective steerability. The security of the protocol remains the same as for entanglement based QKD schemes~\cite{E91,Sec_qkd, Crypt_review}. QKA also offers various advantages over QKD and is manifested in the example we provide in the supplemental material \cite{supp}, where it is used in conjunction with quantum private comparison (QPC) protocols~\cite{qpc1, qpc_epr, qpc_chen}. 

In the next section, we discuss about collective steering where Alice is only stterable by Bob and Charlie together. Using this collective steering we provide a quantum ket authentication protocol in section III. Finally we conclude in section IV.

\section{Collective steering}

 Since we are solely interested in a subset of states for which Alice cannot be steered individually by Bob or Charlie but only by their joint efforts in a tripartite scenario, we start with  a tripartite state $\rho_{abc}$ prepared by Bob (or Charlie). Bob sends the subsystem $A$ to Alice and $C$ to Charlie. Since Alice does not believe Bob or Charlie, she asks them to perform a set of measurements and send her the outcomes. Based on the measurement outcomes, she computes the coherence of her conditional states. We show that there exist states $\rho_{abc}$ for which Alice is steerable if and only if Bob and Charlie make an effort together but not otherwise. To find such a set of states $\{S(A\leftarrow B:C)\}$, we first find out a set of states $\{S(A\leftarrow B, C)\}$ for which Alice is steerable by Bob and Charlie together as well as individually. We then compute the union of set of states $\{S(A\leftarrow B)\}\cup \{S(A\leftarrow C)\}$ for which Alice is steerable by Bob and Charlie individually. Our set of interest is the difference of the above two sets, i.e., $S_{i}\equiv S(A\leftarrow B, C)\setminus \{S(A\leftarrow B)\cup S(A\leftarrow C)\}$.  

We now explicitly find out a set of collectively steerable states. First, we focus on to single out the first set $\{S(A\leftarrow B, C)\}$ (set $I$). Alice will be convinced that her state is entangled if her system $A$ cannot be written by a local hidden state (LHS) model
\begin{equation}\label{LHS model}
\rho_{bc}^{BC}=\sum_{\lambda}\mathcal{P}(\lambda)\mathcal{P}(b,c|B,C,\lambda)\rho_A^Q(\lambda),
\end{equation}
where $\{\mathcal{P}(\lambda),\rho_A^{Q}(\lambda)\}$ represents an ensemble of pre-existing local hidden states of Alice and $\mathcal{P}(b,c|B,C,\lambda)$ is Bob and Charlie's joint stochastic map to convince Alice by preparing a state $\rho_{bc}^{BC}$. $\mathcal{P}(\lambda)$ forms a valid probability distribution such that $\sum_{\lambda}\mathcal{P}(\lambda)=1$. The joint probability distribution on such states can be written as,
\begin{equation}\label{joint probability distribution}
\mathcal{P}(a_{i},b_{j},c_{k})=
\sum_{\lambda}\mathcal{P}(\lambda)\mathcal{P}(b_{j},c_{k}|\lambda)\mathcal{P}^Q(a_i|\lambda),
\end{equation}
where $\mathcal{P}(a_{i},b_{j},c_{k})$ represents the probability to obtain outcome $a$ for the measurement of observables chosen from the set $\{\mathcal{A}_i\}$ by Alice, outcome $b$ for the measurement of observables chosen from the set $\{\mathcal{B}_j\}$ by Bob and outcome $c$ for the measurement of observables chosen from the set $\{\mathcal{C}_k\}$ by Charlie.

We consider a tripartite state $\rho_{abc}$ distributed between Alice $(A)$, Bob $(B)$ and Charlie $(C)$. Alice asks Bob and Charlie to perform projective measurements on their respective systems $(B)$ and $(C)$ on stated bases. We consider Bob and Charlie to perform projective measurements in Pauli eigenbases (or in general on a set of mutually unbiased bases) and communicate the results to Alice. Upon receiving the results, Alice measures coherences on her conditional states with respect to her Pauli eigenbases (or a mutually unbiased bases) with the choice of basis being based on the measurement results from Bob and Charlie. It can be seen that Bob, together with Charlie can steer the state of Alice if at least one steering inequality is violated. Below we provide two such local unitary equivalent inequalities (see supplemental material \cite{supp} for more inequalities)
\begin{eqnarray}
&&\sum_{i\neq k,j,b,c}p(\rho_{A|\Pi^b_i\Pi^c_j})C_k(\rho_{A|\Pi^b_i\Pi^c_j})\leqslant 6\epsilon, \label{steering criteria with charlie 1}\\
&&\sum_{i=j=k,b,c}p(\rho_{A|\Pi^b_i\Pi^c_j})C_k(\rho_{A|\Pi^b_i\Pi^c_j})\leqslant \epsilon, \label{steering criteria with charlie 3}
\end{eqnarray}
Unless otherwise stated we consider $\epsilon=\sqrt{6}$ as shown in~\cite{deba1,deba2} for $l_1$-norm measure of quantum coherence $C_{i}(\rho)$ of the state $\rho$ with respect to the basis $i$, $\{i, j, k\}\in\{0, 1, 2\}$ and $\{a, b, c\}\in\{0, 1\}$. 

 Now, to prove the criteria~(\ref{steering criteria with charlie 1}), we consider that the conditional states of Alice have a local hidden state model as given in Eq.~(\ref{LHS model}), i.e., 
$\rho_{A|\Pi^b_i\Pi^c_j}\equiv\frac{\rho^{BC}_{\Pi^b_i\Pi^c_j}}
{p(\rho^{BC}_{\Pi^b_i\Pi^c_j})}$. Thus,
\begin{eqnarray}\label{proof of steering criteria with charlie 1}
&&\sum_{i\neq k,j,b,c}p\Big(\rho^{BC}_{\Pi^b_i\Pi^c_j}\Big)
C_k\bigg(\frac{\rho^{BC}_{\Pi^b_i\Pi^c_j}}
{p(\rho^{BC}_{\Pi^b_i\Pi^c_j})}\bigg)\nonumber\\
&=&\sum_{i\neq k,j,b,c}p\Big(\rho^{BC}_{\Pi^b_i\Pi^c_j}\Big)
C_k\bigg(\frac{\sum_{\lambda}\mathcal{P}(\lambda)\mathcal{P}(b,c|\Pi^b_i\Pi^c_j,\lambda)\rho_A^Q(\lambda)}{p(\rho^{BC}_{\Pi^b_i\Pi^c_j})}\bigg)\nonumber\\
&\leqslant &\sum_{i\neq k,j,b,c,\lambda}p\Big(\rho^{BC}_{\Pi^b_i\Pi^c_j}\Big)
\frac{\mathcal{P}(\lambda)\mathcal{P}(b,c|\Pi^b_i\Pi^c_j,\lambda)}{p(\rho^{BC}_{\Pi^b_i\Pi^c_j})}C_k\Big(\rho_A^Q(\lambda)\Big)
\nonumber\\
&=&\sum_{i\neq k,j,b,c,\lambda}\mathcal{P}(\lambda)\mathcal{P}(b,c|\Pi^b_i\Pi^c_j,\lambda)C_k\Big(\rho_A^Q(\lambda)\Big)
\nonumber\\
&=&\sum_{k,j,c,\lambda}2\mathcal{P}(\lambda)\mathcal{P}(c|\Pi^c_j,\lambda)C_k\Big(\rho_A^Q(\lambda)\Big)\nonumber\\
&\leqslant &\sum_{j,c,\lambda}2\mathcal{P}(\lambda)\mathcal{P}(c|\Pi^c_j,\lambda)\epsilon
=6\epsilon\nonumber.
\end{eqnarray}

Now we focus on to single out the second set (set $II$), i.e., $S(A\leftarrow B)\cup S(A\leftarrow C)\}$. This is the union of sets of states for which Alice $(A)$ can be steered individually by Bob $(B)$ and Charlie $(C)$. In this case, Alice ignores the results sent by one party while acknowledging the other. A set of steering inequalities in this two-qubit scenario, where Alice ignores the results of Charlie, can be constructed as~\cite{deba1,deba2}
\begin{eqnarray}
&&\sum_{i=k,b}p(\rho_{A|\Pi^b_i})C_k(\rho_{A|\Pi^b_i})\leqslant \epsilon, \label{steering criteria without charlie 1}\\
&&\sum_{i\neq k,b}p(\rho_{A|\Pi^b_i})C_k(\rho_{A|\Pi^b_i})\leqslant 2\epsilon \label{steering criteria without charlie 2}
\end{eqnarray}
and similarly, when Alice ignores the results of Bob, can be expressed as
\begin{eqnarray}
&&\sum_{j=k,c}p(\rho_{A|\Pi^c_j})C_k(\rho_{A|\Pi^c_j})\leqslant \epsilon, \;\mbox{and} \label{steering criteria without alice 1}\\
&&\sum_{j\neq k,c}p(\rho_{A|\Pi^c_j})C_k(\rho_{A|\Pi^c_j})\leqslant 2\epsilon. \label{steering criteria without alice 2}
\end{eqnarray}
We denote all local unitary equivalent inequalities of Eqs.~(\ref{steering criteria with charlie 3}) as the first set and Eqs.~(\ref{steering criteria without charlie 1})-(\ref{steering criteria without alice 2}) as the second set of inequalities. It is our aim to look for a set of states which violate at least one of the first set but not the second set of inequalities. This would ensure that the state of Alice can only be steered by Bob and Charlie together but not individually. 

It is well known fact that a state is deemed steerable if a steering inequality is violated. However, the converse is not always true. 
Thus, there is no definite way to single out the set of such unsteerable states as is required in the 2-qubit scenario to define the set of our interest $(S_i)$. To overcome this issue, we need to use the tightest steering inequalities with semi-definite programming and the free will to choose the bases. 
However, one may start with the set of states for which bi-partite entanglements i.e., $E_{AB}$ and $E_{AC}$ are zero. For such states, by definition, Alice cannot be steered in the 2-qubit scenarios.

For example, we consider a genuine entangled state such as a generalized GHZ state  
\begin{equation}\label{generalized GHZ state}
\ket{\psi}=\alpha\ket{000}+\sqrt{1-\alpha^2}\ket{111},
\end{equation}
where $0\leq\alpha\leq 1$. For the state, it can be shown that no inequality from the second set is violated. This is due to the fact that the entanglement between Alice-Bob $(E_{AB})$ and Alice-Charlie $(E_{AC})$ are zero for GHZ states. On the other hand, the inequality in Eq.(\ref{steering criteria with charlie 3}) from the first set is violated for a certain range of $\alpha$. 

In this sense, a non-trivial example would be $W$ state, i.e., $\ket{\psi}_W=\frac{1}{\sqrt{3}}(\ket{001}+\ket{010}+\ket{100})$, for which $(E_{AB})$ and $(E_{AC})$ are non-zero (see supplemental material \cite{supp}).

\begin{figure}[t]
\centering
\includegraphics[scale=0.7]{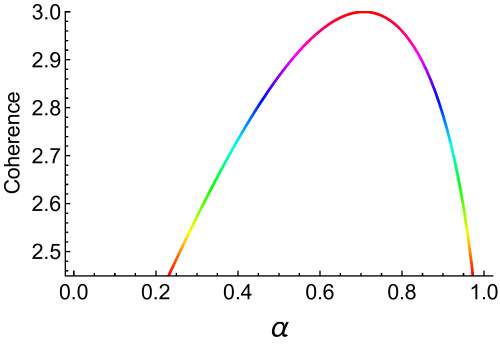}
\caption{Plot of left hand side of Eq. (\ref{steering criteria with charlie 3}) considering Pauli bases with respect to arbitrary reference frame (see supplemental material \cite{supp}) vs $\alpha$ for generalized GHZ state in Eq. (\ref{generalized GHZ state}). For this range of $\alpha$ under the curve, inequality (\ref{steering criteria with charlie 3}) shows violation.}
\label{generalized ghz state coherence}
\end{figure}

However, in the following sections, we will only be concerned about the maximal violation of the steering inequality (\ref{steering criteria with charlie 3}) for our quantum key authentication protocol (QKA). We consider the same set of measurement bases for Alice, Bob and Charlie. They follow certain protocols to choose a basis as depicted by these inequalities. For each inequality or protocol and a fixed set of measurement bases, there must be an unique state violating the inequality maximally. For example, the state (\ref{generalized GHZ state}) shows the maximal violation at $\alpha=\frac{1}{\sqrt{2}}$ only for the inequality (\ref{steering criteria with charlie 3}) as shown in Fig. (\ref{generalized ghz state coherence}). 

\section{Quantum Key Authentication Protocol}

 The non-local features of collectively steerable states can be exploited in many information processing scenarios where a collective action is required. These include but are not limited to reference frame alignment for key distribution~\cite{ref_frame1,ref_frame2,ref_frame3}, secret sharing~\cite{secret_sharing}, escrow services, banking and applications in QPC. We briefly explore some of the scenarios below 
%

Let us consider a scenario in which Bob and Charlie wish to share a secure key. They have fixed a protocol for the same, which includes performing projective measurements in various pre-determined bases. At the end of the protocol, both the parties disclose the choice of basis of measurements but not the results which are kept as part of the key. They keep only those results as key for which their choice of basis matched. However, it may be the case that one of the parties claims to have performed a measurement in a particular basis, when in reality (s)he has not due to misalignment of reference frames or to swindle another. This issue is more relevant when their labs are far apart and there is no way to validate their reference frames. Collective quantum steering provides us with a simple solution to this problem. Instead of sharing a bipartite maximally entangled states (as is usually the case for two party entanglement based QKD), Bob and Charlie share a tripartite $GHZ$ state, such that one of the parties keeps with himself two subsystems instead of one. Without loss of generality, we assume that Charlie prepares a $GHZ$ state and transmits one of the particles to Bob, while he keeps two with himself. This way they share an ensemble of $GHZ$ states.

To align their reference frames, Bob and Charlie perform Pauli measurements on their respective subsystems as proposed by the protocol in Eq.~(\ref{steering criteria with charlie 3}), while one of the subsystems is kept undisturbed by Charlie to check for steering of coherence. The $GHZ$ state exhibits maximal violation of the coherence complimentary relation~\cite{supp} and a local unitary rotation of their bases is needed to perfectly align their reference frames~\cite{supp}. Even in the case when Bob tries to cheat Charlie by not performing the claimed measurements, steerability of quantum coherence can be used to verify the authenticity. A maximum violation of the steering inequality~(\ref{steering criteria with charlie 3}) implies that Bob has indeed performed a measurement in the claimed basis.

In a second possible scenario, a third party Alice provides an escow service to two parties Bob and Charlie. 

Bob and Charlie randomly decide on one of the steering inequalities given by Eq.~(\ref{steering criteria with charlie 1},\ref{steering criteria with charlie 3}) or other inequalities listed in Appendix $B$ and their local unitarily equivalent ones. For each of the inequalities, there exists a unique state which violates it maximally. Alice prepares the requisite state and distributes it among Bob, Charlie and herself. Bob and Charlie then perform key distribution among themselves. Alice authenticates the claims of either party by measuring the coherence of her state with respect to bases as directed by either Bob or Charlie. If the corresponding inequality is violated maximally, Alice is assured that the parties have performed the claimed measurements on their subsystems and the key is likewise authenticated. We present an example of the protocol in which Bob and Charlie decide on the steering inequality~(\ref{steering criteria with charlie 3}) with the $GHZ$ state showing a maximal violation.

Alice prepares the three qubit $GHZ$ state, 
\begin{equation}
|\psi\rangle = \frac{1}{\sqrt{2}}(|000\rangle + |111\rangle),
\end{equation}
which is steerable maximally by Bob and Charlie together when they are measuring in the same Pauli bases but not individually as shown in Fig.~(\ref{generalized ghz state coherence}). Alice distributes this state among Bob, Charlie and herself. Bob and Charlie perform projective measurements $\Pi_i(\theta)$ (see supplemental material~\cite{supp}) on their subsystems and disclose their choice of basis exactly like in standard key distribution schemes. Alice then authenticates whether either party has indeed performed the measurements as claimed by measuring the coherence on her subsystems and plugging in the various values in the steering inequality in Eq.~(\ref{steering criteria with charlie 3}). 

If Bob and Charlie had followed the protocol as described by the steering inequality given in Eq. (\ref{steering criteria with charlie 3}) and used the general measurements as shown in~\cite{supp}, the coherence of the state of Alice would have been steered and the inequality in Eq.~(\ref{steering criteria with charlie 3}) violated maximally. However, if at least one of them had performed the measurements in a different set of bases, the inequality would not be maximally violated. Thus, Alice can validate the key based on the violation of the steering inequality in Eq.~(\ref{steering criteria with charlie 3}) without having any knowledge about the key. 

As a standalone QKD protocol, it falls in the category of entanglement based schemes, for which the minimum secure key rate can be evaluated as follows. Since the parties share a non-local state, their measurement outcomes are correlated to each other which we denote by $\bar{\Pi}$. For the specific example of a $GHZ$ state shared between the parties, it is seen that $\bar{\Pi} = \Pi^b_i\Pi^c_j$. The secure key rate of the protocol is defined as
	
\begin{equation}
	r_{min}\geq I(\Pi_k:\bar{\Pi})-\chi(\Pi_k:E),
	\label{eq: secure_key}
	\end{equation}

where we recall that $\Pi_k$ is the measurement basis chosen by Alice while $E$ is the measurement performed by an eavesdropper Eve and $I(X:Y)$ denotes the mutual information between the variables $X$ and $Y$. The minimization is taken over all the measurement basis being employed by Alice in the protocol. In terms of conditional entropies, Eq.~(\ref{eq: secure_key}) reduces to,
	
\begin{equation}
	r_{min}\geq S(\Pi_k|E)-H(\Pi_k|\bar{\Pi}),
	\label{eq: secure_key2}
	\end{equation}

where, $S(X|Y)$ and $H(X|Y)$ denote the von-Neuman and Shannon entropy of $X$ conditioned on $Y$ respectively. From a result in~\cite{uncertainty}, Eve's uncertainty on the key can be lower bounded as
	
\begin{equation}
	S(\Pi_k|E)+S(\Pi_{k'}|\bar{\Pi})\geq \log_2\left(\frac{1}{v}\right),
	\label{eq: uncertainty}
	\end{equation}

where, $v:=\underset{k,k'}{\text{max}}|\langle\psi_k|\phi_{k'}\rangle|^2$ and $|\psi_k\rangle$, $|\phi_{k'}\rangle$ are eigenvectors of of the measurement basis $\Pi_k$ and $\Pi_{k'}$ respectively. For the specific case of our protocol where Alice uses mutually unbiased Pauli bases for a single qubit, we have $v=2$. Substituting (\ref{eq: uncertainty}) in (\ref{eq: secure_key2}), we get,
	
\begin{equation}
	r_{\text{min}} \geq 1-H(\Pi_k|\bar{\Pi})-H(\Pi_{k'}|\bar{\Pi}).
	\label{eq: secure_key3}
	\end{equation}

For the example of $GHZ$ states with correlations defined as above, the secure key rate is found to be $r_{\text{min}}\geq 1$.








It should be noted that Bob and Charlie will be able to perform key distribution for any state of the form given in Eq. (\ref{generalized GHZ state}) with unit key rate when measurements are performed in the same basis. Similarly, for all such states, the keys can be authenticated using our protocol with certainty as long as the maximum possible violation for the state is known and the state shows exactly the same violation of the steering inequality. For the cases when a maximum violation is not observed for the particular state, the keys distributed between Bob and Charlie cannot be authenticated. Furthermore, our protocol also offers an advantage in reference frame alignment which forms one of the main ingredients in QKD. Although, the aforementioned protocol falls in the category of entanglement based QKD, it is quite different from the traditional ones. The protocol is semi-device independent, in the sense that two interested parties perform key distribution with the help of a third party who authenticates it. The key rate and security analysis performed above are similar to the ones presented in~\cite{secret_sharing}. However, unlike other standard semi-device independent protocols, the aim of our protocol is to authenticate the claims (of measurement) of either party.

\section{Conclusion}

In this letter, we exploit the asymmetric nature of quantum steering for discrete systems and find the set of collectively steerable states for $3$ qubit systems. We find that $GHZ$ and $W$ states are good example of such states. This novel concept leads to new interesting applications in quantum cryptography and communications particularly, in protocols which need concerted action and consensus among parties. We provide one such application in the form of a QKA protocol. By the application of steering of coherence, QKA enables the parties to perfectly align their reference frames, which is a resource intensive task classically. Furthermore, QKA proves helpful in detecting dishonest parties in quantum key distribution. Many contemporary key distribution scenarios deal with third parties; examples include a customer buying a product on an online web-portal. The web-portal hosts various distributors and upon a purchase, authenticates the transaction between customers and the distributors. Our scheme functions in a similar regard, albeit in a more secure manner. Furthermore, it can also be used in conjunction with other information processing tasks such as QPC \cite{supp}. 

We showed that states exhibiting the collective steerability prove advantageous in tasks which are classically hard to achieve. It will be interesting in the future to explore and develop more new applications of such states. We believe, exploring resource theoretic properties of such states and using such states as resource would provide a deeper understanding of the nature of quantum correlations and non-locality altogether. 

\section*{Acknowledgements}

D.M. would like to acknowledge the support from the
National Research Foundation. D.K. is supported by the National
Research Foundation and the Ministry of Education
in Singapore through the Tier 3 MOE2012-T3-1-009 Grant
“Random numbers from quantum processes.”
\bibliography{steering}

\section*{Appendix}

\subsection{Local unitary equivalent steering inequalities}
In Eq. (4), we have provided one of the simplest form of steering inequalities for the tripartite scenario. Other local unitary equivalent steering inequalities based on the Pauli bases can be written as
\begin{eqnarray}
&&\sum_{i,j\neq k,b,c}p(\rho_{A|\Pi^b_i\Pi^c_j})C_k(\rho_{A|\Pi^b_i\Pi^c_j})\leqslant 6\epsilon, \label{steering criteria with charlie 2}\\
&&\sum_{i=j\neq k,b,c}p(\rho_{A|\Pi^b_i\Pi^c_j})C_k(\rho_{A|\Pi^b_i\Pi^c_j})\leqslant 2\epsilon, \label{steering criteria with charlie 4}\\
&&\sum_{i\neq j=k,b,c}p(\rho_{A|\Pi^b_i\Pi^c_j})C_k(\rho_{A|\Pi^b_i\Pi^c_j})\leqslant 2\epsilon\; \mbox{and} \label{steering criteria with charlie 5}\\
&&\sum_{i=k\neq j,b,c}p(\rho_{A|\Pi^b_i\Pi^c_j})C_k(\rho_{A|\Pi^b_i\Pi^c_j})\leqslant 2\epsilon. \label{steering criteria with charlie 6}
\end{eqnarray}
It is not difficult to guess the other local unitary equivalent inequalities based on the permutations and combinations of Pauli bases. One can also use arbitrary mutually unbiased bases (free will) to form the steering inequalities.

\subsection{Example of states with zero bi-partite entanglements $E_{AB}$ and $E_{AC}$ }

Here, we first define Pauli matrices in an arbitrary $\bf{\Theta}$ direction such that the eigenvectors for $\sigma_z(\bf{\Theta})$ are $\ket{z^{+}(\bf{\Theta})}=$ $ \begin{pmatrix} \cos\frac{\theta}{2} \\ e^{-i\phi}\sin\frac{\theta}{2} \end{pmatrix} $ and $\ket{z^{-}(\bf{\Theta})}=$ 
$ \begin{pmatrix} \sin\frac{\theta}{2} \\ -e^{-i\phi}\cos\frac{\theta}{2} \end{pmatrix} $, the eigenvectors for $\sigma_x(\bf{\Theta})$ are $\ket{x^{+}(\bf{\Theta})}=\frac{\ket{z^{+}(\bf{\Theta})}+\ket{z^{-}(\bf{\Theta})}}{\sqrt{2}}$ and $\ket{x^{-}(\bf{\Theta})}=\frac{\ket{z^{+}(\bf{\Theta})}-\ket{z^{-}(\bf{\Theta})}}{\sqrt{2}}$ and the eigenvectors for $\sigma_y(\bf{\Theta})$ are $\ket{y^{+}(\bf{\Theta})}=\frac{\ket{z^{+}(\bf{\Theta})}+i\ket{z^{-}(\bf{\Theta})}}{\sqrt{2}}$ and $\ket{y^{-}(\bf{\Theta})}=\frac{\ket{z^{+}(\bf{\Theta})}-i\ket{z^{-}(\bf{\Theta})}}{\sqrt{2}}$. We consider that Alice, Bob and Charlie--- all start with the Pauli operators and perform measurements on the bases of Pauli operators but the reference frames of Bob and Charlie are aligned at an angle $\bf{\Theta}\equiv(\theta, \phi)$ with respect to that of Alice.

 Suppose Bob and Charlie measuring in $\theta$, $\phi$ direction. Inequality in Eq. $(1)$
 shows maximum violation for the state $\frac{1}{\sqrt{2}}(\ket{000}+\ket{110})$ measured in the basis with $\theta=\pi$ and $\phi=\pi$. The state $\frac{1}{\sqrt{2}}(\ket{010}+\ket{111})$ with $\theta=\pi$ and $\phi=\pi$ violates 
 the inequality in Eq. $(2)$ maximally. Inequality in Eq. $(3)$
 is violated maximally by the state $\frac{1}{\sqrt{2}}(\ket{000}+\ket{111})$ for the measurement with $\theta=\frac{\pi}{2}$ and $\phi=0$. Inequality in Eq. $(4)$
is violated maximally by the state $\frac{1}{2}(\ket{010}+\ket{011}+\ket{110}+\ket{111})$ for $\theta=\frac{\pi}{2}$, $\phi=\pi$. The next inequality in Eq. $(5)$
 shows maximum violation for the state $\frac{1}{\sqrt{2}}(\ket{000}+\ket{110})$ and $\theta=\frac{\pi}{2}$, $\phi=\frac{3\pi}{2}$. For the state $\frac{1}{2}(\ket{010}+\ket{011}+\ket{110}+\ket{111})$ inequality in Eq. $(6)$ is maximally violated for $\theta=0$ and $\phi=\frac{3\pi}{2}$. 
 
 \subsection{Example of states with non-zero bi-partite entanglements $E_{AB}$ and $E_{AC}$ }
 
Let us take a generalized $W$ state of the form $\ket{\psi}_{GW}=\frac{1}{5}\ket{001}+\sqrt{\frac{3}{5}}\ket{010}+\frac{3}{5}\ket{100}$. This state gives violation for set $I$ and set $II$ both. In set $I$ first inequality gives a value of $17.4464$, which is greater than $6\sqrt{6}\approx 14.6969$. Second one gives $14.5289$ which is less than $6\sqrt{6}\approx 14.6969$. Third one is violated by $2.92952$ which is greater than $\sqrt{6}\approx 2.4495$. Fourth one give $5.79661$ which is more than $2\sqrt{6}\approx 4.8989$. Fifth and sixth one give $5.88952$ which is also more than $2\sqrt{6}\approx 4.8989$. Now we will see what about the second set for $W$ state. First one give $2.82029$, which is greater than $\sqrt{6}\approx 2.4495$. For second one we get $5.64058$ which is more than $2\sqrt{6}\approx 4.8989$. Third one give $0.893575$ which is less than $\sqrt{6}\approx 2.4495$. The final one give $1.77809$ which is again less than $2\sqrt{6}\approx 4.8989$. 

On the other hand, let us take the three qubit $W$ state $\ket{\psi}_W=\frac{1}{\sqrt{3}}(\ket{001}+\ket{010}+\ket{100})$. For this state inequalities of the set $I$ are violated but not of set $II$. In set $I$ first inequality gives a value of $15.6835$, which is greater than $6\sqrt{6}\approx 14.6969$. Second one gives $15.6835$ which is also greater than $6\sqrt{6}\approx 14.6969$. Third one is violated by $2.86603$ which is greater than $\sqrt{6}\approx 2.4495$. Fourth one give $5.4664$ which is more than $2\sqrt{6}\approx 4.8989$. Fifth and sixth one give $5.69936$ which is also more than $2\sqrt{6}\approx 4.8989$. Now we will see what about the second set for $W$ state. After tracing out Charlie the entanglement of the reduced density matrix of Alice and Bob is $\frac{2}{3}$. First one give $1.84424$, which is less than $\sqrt{6}\approx 2.4495$. For second one we get $3.54606$ which is again less than $2\sqrt{6}\approx 4.8989$. Now after tracing out Bob the concurrence of reduced density matrix of Alice and Charlie is again $\frac{2}{3}$. Third one give $1.84424$ which is less than $\sqrt{6}\approx 2.4495$. The final one give $3.54606$ which is again less than $2\sqrt{6}\approx 4.8989$.  

\subsection{Authentication protocol in conjunction with QPC protocol}

Since our scheme utilizes $GHZ$ states, it provides an added advantage in that existing QPC protocols like Chen's protocol ~\cite{qpc_chen} can be easily modified. In Chen {\it et.} {\it al.}'s protocol, Alice prepares a sequence of $GHZ$ states which she distributes among Bob, Charlie and herself. After performing a privacy amplification of the quantum channel, Bob and Charlie perform a projective measurement in the $\sigma_x$ basis. Bob and Charlie declare the results of their measurements in the following manner: if the result corresponds to their secret message they announce $0$, and $1$ otherwise. Depending upon the announced results, Alice performs a unitary evolution of her subsystem followed by a projective measurement in the $\sigma_x$ basis. The outcome to this measurement reveals the equality of Bob and Charlie's information. No where in the protocol Bob and Charlie announce their information publicly.

In the above protocol it is implicitly assumed that Bob and Charlie's devices are uncorrelated to an eavesdropper. Such a correlation with an eavesdropper corresponds to additional degrees of freedom which in turn implies that the devices can not essentially perform qubit measurements. In order to elevate the protocol to a semi-device independent status, an additional check in form of QKA can be implemented prior to the protocol. This single check serves two purposes: firstly, it can help to verify the security of the quantum channel to be used and secondly, it can help to authenticate the measurements of Bob and Charlie. Such co-joining of two information processing tasks, although requiring slightly higher number of resources (in the form of $GHZ$ states), yields not one but two advantages over the standard QPC protocols.

In the modified protocol, Alice prepares a long sequence of $GHZ$ states of length $L$ and distributes them amongst Bob, Charlie and herself. Bob and Charlie select a basis from the set of arbitrarily chosen mutually unbiased bases to perform the measurements. The selection made by them is not random, but is rather biased towards a particular basis. For a length $l<L$, which includes all the $\sigma_z$ measurements and a randomly selected equal number of $\sigma_x$ measurements, all the three parties perform QKA as detailed above. For the maximal violation of inequality~(4), the coherence of Alice's state is steered maximally, which in turn implies that Bob and Charlie's devices indeed perform as claimed and they are not cheating. For the remaining sequence of states of length $L-l$, the parties perform QPC remain untouched as detailed in~\cite{qpc_chen}. Thus with the help of QKA, we ensure privacy amplification of the quantum channel as well as elevate the protocol to a semi-device independent status, without the cost of using any additional features.

\end{document}